# Unified System for Processing Real and Simulated Data in the ATLAS Experiment

© Mikhail Borodin
Big Data Laboratory, National Research Centre "Kurchatov Institute",
Moscow, Russia
National Research Nuclear University "MEPhI",
Moscow, Russia
mikhail.borodin@cern.ch

© Kaushik De
University of Texas
Arlington, TX, U.S.A.
kaushik@uta.edu

© Jose Garcia Navarro
IFIC, Universidad de Valencia,
Valencia, Spain
jose.enrique.garcia@cern.ch

© Dmitry Golubkov
Big Data Laboratory, National Research Centre "Kurchatov Institute",
Moscow, Russia
Institute for High Energy Physics,
Protvino, Russia
dmitry.v.golubkov@cern.ch

© Alexei Klimentov
Big Data Laboratory, National Research Centre "Kurchatov Institute",
Moscow, Russia
Brookhaven National Laboratory,
Brookhaven, NY, U.S.A.
aak@bnl.gov

© Tadashi Maeno
Brookhaven National Laboratory,
Brookhaven, NY, U.S.A.
maeno@bnl.gov

© David South
Deutsches Elektronen-Synchrotron,
Hamburg, Germany
southd@mail.desy.de

© Alexandre Vaniachine
Argonne National Laboratory,
Argonne, IL, U.S.A.
vaniachine@anl.gov
on behalf of the ATLAS Collaboration

## Abstract

The physics goals of the next Large Hadron Collider run include high precision tests of the Standard Model and searches for new physics. These goals require detailed comparison of data with computational models simulating the expected data behavior. To highlight the role which modeling and simulation plays in future scientific discovery, we report on use cases and experience with a unified system built to process both real and simulated data of growing volume and variety.

## 1 Introduction

In 2015 the Large Hadron Collider will open new "Gates of Nature" by reaching instantaneous luminosities exceeding $2·10^{34}$ cm$^{-2}$s$^{-1}$ and center of mass energies of 13 TeV. The physics goals of the ATLAS experiment [1] include searches for physics beyond the Standard Model and high precision Higgs sector studies. These goals require detailed comparison of the expected physics and detector behavior with data.

This comparison is missed in the naïve ("let the data speak for themselves") approach to Big Data processing, which omits modeling and simulation. In contrast to the naïve approach, in our case the comparison requires a large number of computational models simulating the expected behavior. In order to simulate the diversity of LHC physics, ATLAS produces more than 35 000 samples of simulated data per year. The variety of our simulated data adds value to the unified system for Big Data processing.

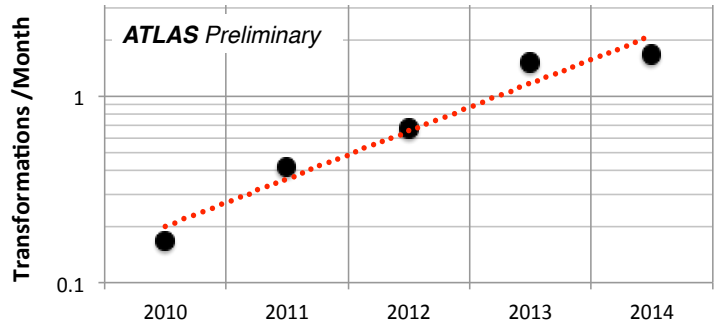

*Figure 1: Continuous growth in the rate of data transformations handled by the production system.*

The LHC experiments employ the computational infrastructure of the Worldwide LHC Computing Grid (WLCG) [2]. The academic distributed computing environment of WLCG builds on the ideas proposed by Foster and Kesselman [3]. ATLAS leads in the WLCG usage in the number of data processing jobs and processed data volume. ATLAS manages more than 160 petabytes of data on more than hundred computational sites.

The central ATLAS production system (here referred to as ProdSys1) was designed and deployed more than nine years ago to address the needs of simulations and real data processing. Since the start of its operation, ProdSys1 proved itself as a valuable tool for processing real and simulated data on the Grid and became one of the enabling factors for the Collaboration.

ATLAS experience gained in managing simulations and data processing on an ever-increasing scale and handling scientific workflow requests of growing complexity led to a realization that ProdSys1 has to be significantly enhanced and updated. In this paper we highlight the production system techniques validated through practice, describe representative use cases handled by the production system and its upgrade – the ProdSys2.

## 2 Experience with Techniques

For data processing The ATLAS experiment adopted the data transformation technique, where software applications transform the input datasets of one data type (described in Appendix) into the output datasets of another data type. Figure 1 shows that the data transformation technique experienced an exponential growth rate in the number of new data transformations handled by the production system.

The ATLAS production system deals with datasets, not individual files. Similarly a task (comprised of many data processing jobs) has become a unit of the workflow in ATLAS data processing. The ATLAS production system therefore serves an important role as a top layer of abstraction responsible for defining jobs in a scalable and automated manner for unified processing of the real and simulated data. Jobs are defined in large collections (tasks), and are formulated to fulfil "task requests". A task is defined based on a data transformation request, and it has a number of attributes, set in accordance with that request, while requests are essentially dictionaries of parameters fed into the system by the scientist using a Web interface or other tools.

The variety of our data and data transformations places a significant burden on scientists to configure and manage the large number of parameters and options. The laborious process of steering the data transformations by providing dictionaries of parameters is manual. The error-prone manual process does not scale to the LHC challenges. To reduce human errors and automate the process of defining millions of jobs, the production system manages configuration parameters and guarantees reproducibility of results. The production system management of the institutional knowledge of the process of tuning and setting up data processing tasks resulted in major gains in efficiency and productivity.

The next step in the lifecycle of the task is its automatic translation into a number of individual job definitions, which sometimes is quite large (tens of thousands). In ProdSys1, individual job definitions are set based on the parent task parameters and remain static for the duration of the whole task execution. Data pertaining to requests, tasks and jobs are persisted in the Oracle RDBMS. Thus, ProdSys1 operation can be described as a sequence of transitions, starting with recording requests, and formulating tasks by processing those requests and then creating a static list of job definitions for each task. Figure 2 shows that the production system experienced double exponential growth of the number of task requests.

Leveraging the underlying job management system PanDA described in Ref. [4], the production system steers ATLAS data processing applications on more than the many hundred thousands of CPU-cores provided by the WLCG. The individual data processing tasks are organized into workflows. During data processing, the system monitors site performance and supports dynamic sharing minimizing the workflow duration. Also, the production system enhances the resilience by managing failures in tasks and/or jobs, bulk job submission depending on success of the scout jobs, automatic e-mail notification, etc.

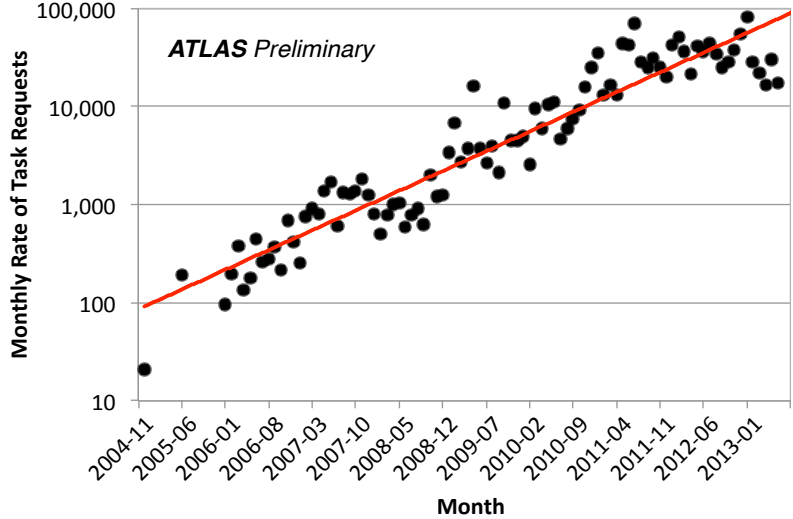

*Figure 2: The monthly rate of tasks requests grows exponentially.*

As the production workload increased in volume and complexity (the ATLAS production tasks count is above 1.6 million), the ATLAS experiment scaled up its data processing capabilities by upgrading the production system that unifies processing of the real and simulated data. To reduce the costs, the production system core remained shared by all use cases, despite the differences in their data processing requirements described in the next section.

## 3 Use Cases and Results

Table 1 lists representative data processing use cases described in the following sub-subsections. The key difference in data processing requirements for the real and simulated data is tolerance to data losses that could happen during processing. The simulated data samples tolerate data loss that reduces the sample statistics without introducing physics bias. In contrast, data losses are not tolerated in real and trigger data processing.

During processing, data losses can occur for a variety of reasons, which are hard to diagnose and repair quickly. The technique of automatic resubmission of the failed jobs recovers data losses from transient failures at the expense of CPU time used by the failed jobs. This technique has not presented a problem, as the fraction of CPU-time used for the recovery of lost data varied from three to six percent, while reducing data losses below the $10^{-8}$ level. The achieved level enables recovery of the lost data on a case-by-case basis.

### 3.1 Trigger Data Processing

The first step to new physics discoveries and high precision studies of rare events is the rejection of the benign events with "known" processes. The multi-tier trigger system reduces the exascale data flood to a petascale level by rejecting data from most of LHC collision events. In 2015, the ATLAS experiment will have a two-tier trigger system:

1. The hardware-based Level 1 trigger.
2. The software-based High-Level Trigger designed by the ATLAS Collaboration [5].

The trigger data processing happens one step before the recording of "raw" data from the ATLAS detector. Thus, any inefficiencies or mistakes may lead to unrecoverable loss of real data. To eliminate such losses, the dedicated trigger data processing workflow is employed to validate trigger software and other critical trigger changes during data taking. The trigger data processing is the main tool for commissioning the trigger for data taking. The challenge here is to achieve the fast turnaround, while avoiding data loss. The experience with ProdSys1 shows that resubmission of the failed jobs delays the task completion. The resubmission technique has to be improved to meet the trigger data processing requirements.

*Table 1: Data processing use cases*

| Use Case | Frequency | Workflow Length | Number of Tasks | Tasks Duration | Data Loss |
|---|---|---|---|---|---|
| Trigger Data | Weekly | Short | Several | Day | no |
| Real Data | Yearly | Medium | Hundreds | Weeks | no |
| Simulated Data | Quarterly | Long | Thousands | Months | yes |

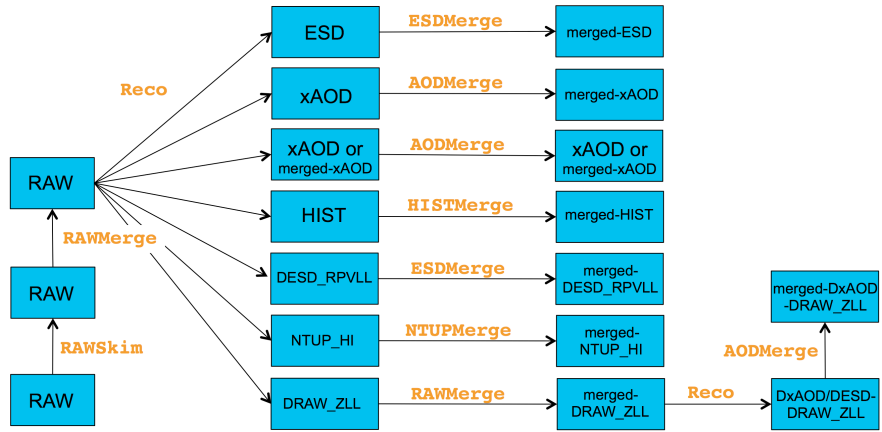

*Figure 3: Data processing workflow for the real data. Arrows are labeled with data transformation applications; boxes are labeled with various data types produced (see Appendix for data types descriptions).*

### 3.2 Real Data Processing

The "raw" data from the ATLAS detector are processed to produce the reconstructed data for physics analysis. During reconstruction ATLAS applications process raw detector data to identify and reconstruct physics objects such as leptons. Figure 3 shows the data processing flow used in reconstruction.

The ATLAS collaboration completed four petascale data processing campaigns on the Grid, with up to 2 PB of real data being processed every year. Table 2 lists parameters for the ATLAS yearly data processing campaigns. (In 2013 reprocessing, 2.2 PB of input data were used for selecting about 15% of all events for reconstruction, thus reducing CPU resources vs. the 2012 reprocessing.)

Thanks to the quality of the ATLAS software and improvements in data processing workflow, in the 2011 data reprocessing only two collision events (out of 0.9 $10^9$ events total) could not be reconstructed due to subtle software bugs. Correcting these bugs allowed the remaining events to be processed in a dedicated data recovery step.

In rare cases, the data corruption during job execution has been missed by the software or computing framework. This is called silent data corruption, which is detected later during data analysis. For example, the silent data corruption was detected in six events from the reprocessed 2010 data and in five adjacent events from the 2011 reprocessed data. Lowering the event losses below the $10^{-8}$ level enables recovery of the lost data on a case-by-case basis.

### 3.3 Simulated Data Processing

The computational resources required to process the simulated data dominate the overall resource usage. The data processing campaigns for the simulated data correspond to the data taking periods of the real data. The LHC data taking periods of the same conditions are characterized by the same center-of-mass energy, instantaneous luminosity, detector configuration, etc.

*Table 2: Processing campaigns for real data*

| Campaign year | Input Data Volume (PB) | CPU Time Used for Reconstruction ($10^6 h$) | Events Processed ($10^9$) | Events not Processed | Silent Data Corruption (events) |
|---|---|---|---|---|---|
| 2010 | 1 | 2.6 | 0.9 | $5 \cdot 10^4$ | 6 |
| 2011 | 1 | 3.1 | 0.9 | 2 | 5 |
| 2012 | 2 | 14.6 | 2.0 | 14 | 0 |
| 2013 | 2 | 4.4 | 0.3 | 4 | 0 |

*Table 3: Data processing campaigns for simulated data*

| Campaign Label | Data Taking Period for Real Data | Configuration | Full Simulation ($10^9$ events) | Fast Simulation ($10^9$ events) | Number of Sub-campaigns |
|---|---|---|---|---|---|
| mc11 | 2011 | 7 TeV | 3.64 | 3.27 | 4 |
| mc12 | 2012 | 8 TeV | 6.37 | 6.43 | 3 |
| mc14 | 2012 & 2015 | 8 & 13 TeV | 0.85 | | 2 |

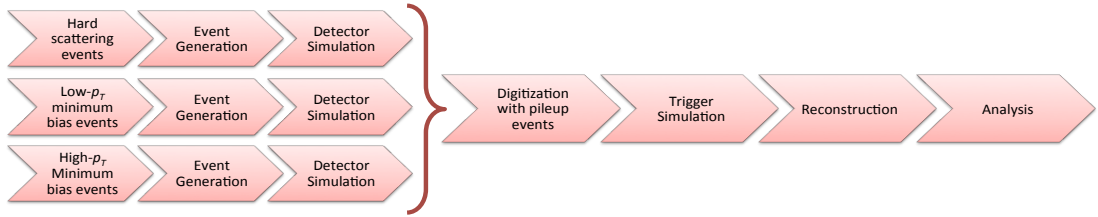

*Figure 4: ATLAS simulations workflow is composed of many steps. Several initial steps are repeated for the hard-scattering events and the minimum bias events (simulated in two complementary samples).*

Table 3 lists the major data processing campaigns for the simulated data.

The LHC instantaneous luminosities result in the presence of a large number of simultaneous collisions in the same event, overlapping the hard scattering event of interest. The presence of these additional minimum bias events is usually called "pileup". To provide realistic simulation of these conditions, the data processing workflow for simulated data is composed of many steps (Figure 4): generate or configure hard-processes, hadronize signal and minimum-bias (pileup) events, simulate energy deposition in the ATLAS detector, digitize electronics response, simulate triggers, reconstruct data, transform the reconstructed data into data types for physics analysis, etc. The intermediate outputs are merged and/or filtered as necessary to optimize the chain.

An example of a more complex workflow used to simulate the ATLAS trigger using dedicated hardware for fast tracking (FTK) designed by the ATLAS Collaboration [6] is shown on Figure 5, where to keep the computational resources for the FTK simulation below practical limits, every event is split into 256 $\eta$-$\phi$ sub-regions [7]. In the three-step workflow, each event is processed by 64 jobs; each job simulates tracks in four FTK sub-regions one after another. The sub-region merging is done in two steps: producing whole regions, then whole events in the NTUP_FTK files. The final step uses FTK tracks in trigger simulations producing the reconstructed data in DESD_FTK files or adds FTK tracks to the simulated events in the RDO_FTK files (see Appendix for the data types descriptions).

The FTK workflow demonstrates the production system workflow with sub-events processing (vs. traditional processing of the whole events). We expect that next-generation processor technologies (such as many-core) will increase the number of workflows with fine-grained processing.

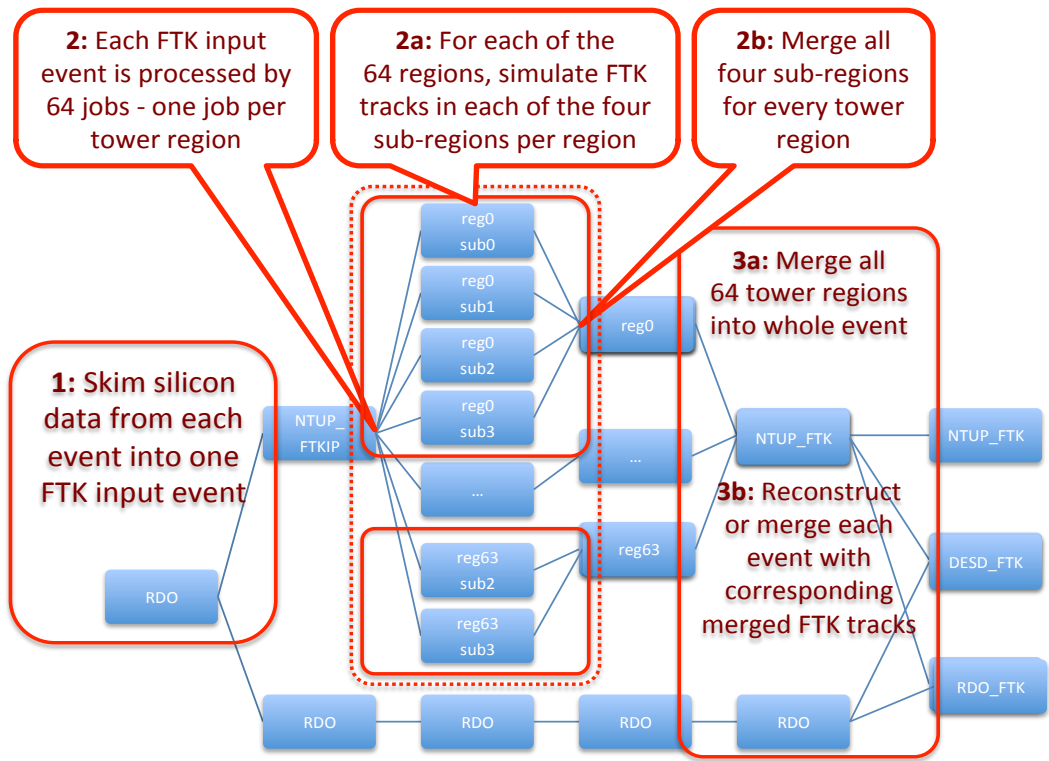

*Figure 5: The simulation of the FTK hardware splits every event into 256 sub-events (see Appendix for descriptions of data types shown in the boxes).*

Validating the Big Data processing techniques, four different sub-campaigns of the mc11 campaign implemented the pileup conditions, detector conditions and geometry increasingly closer to those in real data. During the mc12 campaign, the majority of the events was simulated in the sub-campaign mc12b. Later, the mc12c sub-campaign implemented an improved detector geometry description. The goal of the mc14 campaign was to prepare for the 2015 data taking. The 8 TeV events were processed with improved and updated simulation, digitization and reconstruction software while using the same conditions as in the mc12 campaign. The 13 TeV campaign had the center of mass energy expected for the 2015 data taking with estimated pileup and detector conditions. The mc14 campaign used the new ATLAS Integrated Simulations Framework described in Ref. [8], with multicore processing becoming the default for major simulated data processing steps: simulation, digitization and reconstruction.

## 4 Multilayer Data Processing System

The LHC shutdown provided an opportunity for upgrading the production system, making implementations more maintainable by separating the core concerns: the system logic layer and the presentation layer. Figure 6 shows that on top, the upgraded Task Request interface encapsulates the concise and transparent presentation layer for users, while the lower Task Definition layer implements the core data processing logic that empowers scientists with templated workflow definitions through the Database Engine for Tasks (DEfT). At the layer below, the Job Execution and Definition Interface (JEDI) is integrated with the PanDA layer to provide dynamic job definition tailored to the sites capabilities. The two middle layers communicate via customized JSON protocol. The multi-layer design provided clean separation of the PanDA-specific JEDI layer defining the jobs and the generic workflow abstraction confined to the DEfT layer.

### 4.1 Workflow Abstraction

The ProdSys1 experience made clear that the new level of abstraction – the workflow – is necessary to scale up the system in support of the growing number of tasks and data transformations. The workflow is a group of interdependent tasks, where dependencies exist in the form of the input/output datasets. Such would be the case with a "chain" type simulations workflow (Figure 4), where the data goes through transformations with possible other steps like merging interspersed in between.

DEfT is the service that implements the workflow abstraction in the upgraded production system. Due to relative autonomy of DEfT, i.e. it's agnostic approach to the computational resources used (which are in the scope of the JEDI component), it is appropriate to describe it as a state machine, in which the states of the workflow components undergo a series of transitions based on certain predefined rules and external events, such completion of a job managed by JEDI. By itself, DEfT is not a workload management or resource provisioning system. It is a higher-level abstraction layer encapsulating institutional knowledge that allows scientists to concentrate on the design, monitoring and effective management of production workflows.

### 4.2 Job Definition Improvements

The ProdSys1 experience demonstrated that it is difficult to generalize the procedure of job definition since the method used to define jobs depends on the task type. For improved maintainability and flexibility JEDI has a pluggable structure so that a plug-in defines jobs for each use case.

Dynamic job definition means that the job is defined at some point after the creation of its parent task, and taking into account operating conditions that were not known at the task inception. This is in contrast to a static definition of the job that happens immediately once the task is created. There are advantages realized once there is a capability to define jobs dynamically, based on the actual resources and other conditions present once the task moves into the execution stage. For example, the late binding takes advantage of defining small jobs for the sites with limited capabilities (slow CPU, low disk, network, etc.) or defining extremely large jobs for processing on the HPC facility with hundred thousand CPU-cores. Another advantage is resubmission of the long failed job redefined as several short jobs. This capability avoids delays in task completion for use cases such as trigger data processing described in Section 3.1.

We further upgraded the production system to improve the performance and accommodate a growing number of new requirements and use cases. Figure 7 shows that the upgraded system is deployed in production supporting complex workflows with tasks processing simulated and real data for future scientific discoveries.

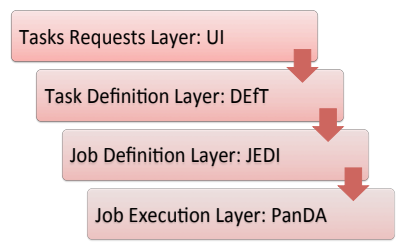

*Figure 6: Multi-layer architecture of the ATLAS production system for processing real and simulated data.*

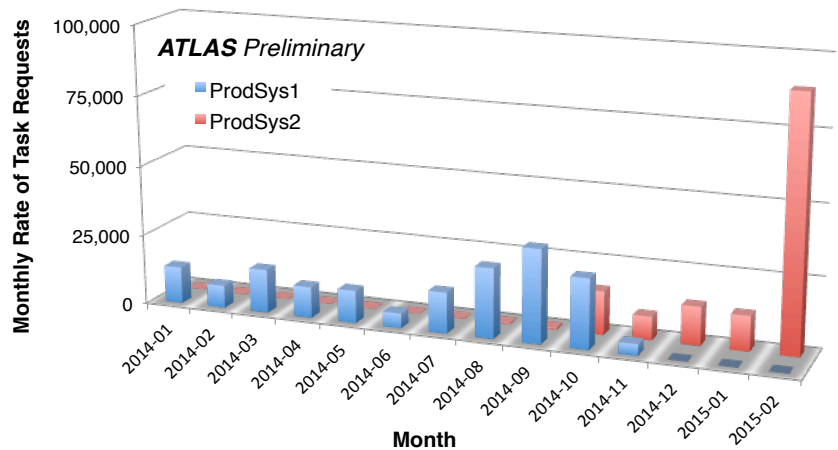

*Figure 7: Since deployment, ProdSys2 exceeded the monthly rate of submitted tasks in ProdSys1.*

## 5 Conclusion

Over the last several years, the ATLAS production system unified a diverse range of workflows and special use cases including processing of the real and the simulated data at large scales. The ATLAS production system fully satisfies the Big Data processing requirements of the ATLAS experiment through the unified approach for real data processing and simulations as well as the mixture of both. This technique enabled to address a much wider range of physics analyses, with sufficiently good accuracy. The unified production system does not limit our ability to process real and simulated data. In addition, detailed physics studies established that the simulated data are of higher accuracy and variety compared to previous generations of experiments, describing the detector behavior quite well in most analyses. The unified capabilities for real and simulated data processing significantly enhanced ATLAS physics output, and they motivated production of higher than foreseen simulated data volumes.

## Acknowledgements

We wish to thank all our colleagues who contributed to ATLAS Big Data processing activities. This work was funded in part by the U. S. Department of Energy, Office of Science, High Energy Physics and ASCR Contract No. DE-AC02-98CH10886 and Contract No. DE-AC02-06CH11357. NRC KI team work was funded by the Russian Ministry of Science and Education under Contract №14.Z50.31.0024.

## Appendix

Table 4 provides abbreviations for the data types used by the ATLAS experiment. During data processing, data reduction is often used to select targeted events and store only the necessary information, taking into account the physics goals and data volume estimates.

*Table 4: Data types used by the ATLAS experiment*

| Short Name | Data Type Name | Description |
| --- | --- | --- |
| RAW | Raw Data | Events selected after the High Level Trigger. |
| ESD | Event Summary Data | Events reconstructed from RAW that contain sufficient information to allow rapid tuning of reconstruction algorithms and detector calibrations. |
| AOD | Analysis Object Data | Contains a summary of the reconstructed event contains sufficient information for common physics analyses. |
| DPD | Derived Physics Data | AOD specific to one or a few analysis groups |
| RDO | Raw Data Object | Representation of the RAW data format used predominantly in simulation. |
| DESD | Derived Event Summary Data | Data derived from the ESD where reduction is used to select targeted events and store only the necessary information. |
| DAOD | Derived Analysis Object Data | Data derived from the AOD where reduction is used to select targeted events and store only the necessary information. |
| NTUP | N-tuples | Contains summary n-tuples for the data processed. |
| HIST | Histograms | Contains summary histograms for the data processed. |